\documentclass[english,12pt]{article}
\usepackage{array}
\usepackage{graphicx}
\usepackage{amssymb}
\usepackage{amsmath}
\usepackage{multirow}
\usepackage{prettyref}
\usepackage{babel}
\usepackage{units}
\usepackage[latin1]{inputenc}
\usepackage{amsfonts}
\usepackage{amssymb}
\usepackage{babel}
\usepackage{cite}
\def\@fmsl@sh#1#2#3{\m@th\ooalign{$\hfil#1\mkern#2/\hfil$\crcr$#1#3$}}
 \def\eq#1\en{\begin{equation}#1\end{equation}}
\def\s[#1,#2]{[#1\stackrel{\star}{,}#2]}
\def\sx[#1,#2]{[#1\stackrel{\star_{x}}{,}#2]}

\newcommand{\nc}{\newcommand}
\nc{\beq}{\begin{equation}}
\nc{\eeq}{\end{equation}}
\nc{\beqa}{\begin{eqnarray}}
\nc{\eeqa}{\end{eqnarray}}

\def\bc{\begin{center}}
\def\ec{\end{center}}

\def\gsim{\mathrel{\mathpalette\atversim>}}

\def\bc{\begin{center}}
\def\ec{\end{center}}

\def\gsim{\mathrel{\rlap{\lower4pt\hbox{\hskip1pt$\sim$}}

    \raise1pt\hbox{$>$}}}       

\def\gsim{\mathrel{\rlap{\lower4pt\hbox{\hskip1pt$\sim$}}
    \raise1pt\hbox{$>$}}}       

\begin{document}
\makeatletter
\def\fmslash{\@ifnextchar[{\fmsl@sh}{\fmsl@sh[0mu]}}
\def\fmsl@sh[#1]#2{%
  \mathchoice
    {\@fmsl@sh\displaystyle{#1}{#2}}%
    {\@fmsl@sh\textstyle{#1}{#2}}%
    {\@fmsl@sh\scriptstyle{#1}{#2}}%
    {\@fmsl@sh\scriptscriptstyle{#1}{#2}}}
\def\@fmsl@sh#1#2#3{\m@th\ooalign{$\hfil#1\mkern#2/\hfil$\crcr$#1#3$}}
\makeatother

\thispagestyle{empty}
\begin{titlepage}
\boldmath
\begin{center}
  \Large {\bf Gravitational Waves in Effective Quantum Gravity}
    \end{center}
\unboldmath
\vspace{0.2cm}
\begin{center}
{  {\large Xavier Calmet}\footnote{x.calmet@sussex.ac.uk}, 
{\large Iber\^ e Kuntz}\footnote{ibere.kuntz@sussex.ac.uk}
and {\large Sonali Mohapatra}\footnote{s.mohapatra@sussex.ac.uk}
}
 \end{center}
\begin{center}
{\sl Physics $\&$ Astronomy, 
University of Sussex,   Falmer, Brighton, BN1 9QH, United Kingdom 
}
\end{center}
\vspace{5cm}
\begin{abstract}
\noindent
In this short paper we investigate quantum gravitational effects on Einstein's equations using effective field theory techniques. We consider the leading order quantum gravitational correction to the wave equation. Besides the usual massless mode, we find a pair of modes with complex masses. These massive particles have a width and could thus lead to a damping of gravitational waves if excited in violent astrophysical processes producing gravitational waves such as  e.g. black hole mergers. We discuss the consequences for  gravitational wave events such as GW 150914 recently observed by the Advanced LIGO collaboration.

\end{abstract}  
\end{titlepage}


The recent discovery of gravitational waves by the Advanced LIGO collaboration \cite{Abbott:2016blz} marks the beginning of a new era in astronomy which could shed some new light on our universe revealing its darkest elements that do not interact with electromagnetic radiations. This discovery could also lead to some new insights in theoretical physics.  In this short paper we study the leading effect of quantum gravity on gravitational waves using effective field theory techniques. While the discovery of a theory of quantum gravity might still be far away, it is possible to use effective field theory techniques to make actual predictions in quantum gravity. Assuming that diffeomorphism invariance is the correct symmetry of quantum gravity at the Planck scale and assuming that we know the field content below the Planck scale, we can write down an effective action for any theory of quantum gravity.  This effective theory, dubbed Effective Quantum Gravity, is valid up to energies close to the Planck mass.  It is obtained by linearizing general relativity around a chosen background. The massless graviton is described by a massless spin 2 tensor which is quantized using the standard quantum field theoretical procedure. It is well known that this theory is non-renormalizable, but divergences can be absorbed into the Wilson coefficients of higher dimensional operators compatible with diffeomorphism invariance. The difference with a standard renormalizable theory resides in the fact that an infinite number of measurements are necessary to determine the action to all orders. Nevertheless, Effective Quantum Gravity enables some predictions which are model independent and which therefore represent true tests of quantum gravity, whatever the underlying theory might be.

We will first investigate quantum gravitational corrections to linearized Einstein's equations. Solving these equations, we show that besides the usual solution that corresponds to the propagation of the massless graviton, there are solutions corresponding to massive degrees of freedom.  If these massive degrees of freedom are excited during violent astrophysical processes a sizable fraction of the energy released by such processes could be emitted into this modes. We shall show that the corresponding gravitational wave is damped and that the energy of the wave could thus dissipate. We then study  whether the recent discovery of gravitational waves by the Advanced LIGO collaboration \cite{Abbott:2016blz} could lead to a test of quantum gravity. 

Given a matter Lagrangian coupled to general relativity with $N_s$ scalar degrees of freedom, $N_f$ fermions and $N_V$ vectors one can calculate the graviton vacuum polarization in the large $N=N_s +3 N_f +12 N_V$ limit with keeping $N G_N$, where $G_N$ is Newton's constant, small. Since we are interested in energies below $M_\star$ which is the energy scale at which the effective theory breaks down, we do not need to consider the graviton self-interactions which are suppressed by powers of $1/N$ in comparison to the matter loops. Note that $M_\star$ is a dynamical quantity and does not necessarily corresponds to the usual reduced Planck mass of order $10^{18}$ GeV (see e.g. \cite{Calmet:2013hfa}). The divergence in this diagram can be isolated using dimensional regularization and absorbed in the coefficient of $R^2$ and $R_{\mu\nu}R^{\mu\nu}$. An infinite series of vacuum polarization diagrams contributing to the graviton propagator can be resummed in the large $N$ limit. This procedure leads to a resummed graviton propagator given by \cite{Aydemir:2012nz}\begin{eqnarray} \label{resprop}
i D^{\alpha \beta,\mu\nu}(q^2)=\frac{i \left (L^{\alpha \mu}L^{\beta \nu}+L^{\alpha \nu}L^{\beta \mu}-L^{\alpha \beta}L^{\mu \nu}\right)}{2q^2\left (1 - \frac{N G_N q^2}{120 \pi} \log \left (-\frac{q^2}{\mu^2} \right) \right)}
\end{eqnarray}
with $L^{\mu\nu}(q)=\eta^{\mu\nu}-q^\mu q^\nu /q^2$ and where $\mu$ is the renormalization scale. This resummed propagator is the source of interesting acausal and non-local effects which have just started to be investigated \cite{Aydemir:2012nz,Donoghue:2014yha,Calmet:2013hia,Calmet:2015dpa,Calmet:2014gya,Calmet:2015pea}. Here we shall focus on how these quantum gravity effects affect gravitational waves.

From the resummed graviton propagator in momentum space, we can directly read off the classical field equation for the spin 2 gravitational wave in momentum space
\begin{eqnarray}
2q^2\left (1 - \frac{N G_N q^2}{120 \pi} \log \left (-\frac{q^2}{\mu^2} \right) \right)=0.
\end{eqnarray}
This equation has three solutions \cite{Calmet:2014gya}:
\begin{eqnarray}
q^2_{1}&=0,\\ \nonumber
q^2_{2}&=& \frac{1}{G_N N} \frac{120 \pi}{ W\left (\frac{-120 \pi}{\mu^2 N G_N} \right)}, \\ \nonumber
q^2_{3}&=&(q^2_{2})^*,
\end{eqnarray}
where $W$ is the Lambert function. The complex pole corresponds to a new massive degree of freedom with a complex mass (i.e. they have a width  \cite{Calmet:2014gya}). The general wave solution is thus of the form
\begin{eqnarray}
h^{\mu\nu}(x)=a^{\mu\nu}_1 \exp(-i q_{1\alpha} x^\alpha)+a^{\mu\nu}_2 \exp(-i q_{2\alpha} x^\alpha)+a^{\mu\nu}_3 \exp(-i q^\star_{2\alpha} x^\alpha).
\end{eqnarray}
We therefore have three degrees of freedom which can be excited in gravitational processes leading to the emission of gravitational waves. Note that our solution is linear, non-linearities in gravitational waves (see e.g. \cite{Aldrovandi:2008ci}) have been investigated and are as expected very small. 

The position of the complex pole depends on the number of fields in the model. In the standard model of particle physics, one has $N_s = 4$, $N_f = 45$, and $N_V = 12$. We thus find $N=283$ and the pair of complex poles at $(7 -  3 i)\times 10^{18}$ GeV and  $(7 + 3 i)\times 10^{18}$ GeV. Note that the pole $q^2_{3}$ corresponds to a particle which has an incorrect sign between the squared mass and the width term. We shall not investigate this Lee-Wick pole further and assume that this potential problem is cured by strong gravitational interactions. The renormalization scale needs to be adjusted to match the number of particles included in the model. Indeed, to a good approximation the real part of the complex pole is of the order of 
\begin{eqnarray}
| \mbox{Re} \ q_2|\sim \sqrt{\frac{120 \pi}{ N G_N}}
 \end{eqnarray}
 which corresponds to the energy scale $M_\star$  at which the effective theory breaks down. Indeed, the complex pole will lead to acausal effects and it is thus a signal of strong quantum gravitational effects which cannot be described within the realm of the effective theory. We should thus pick our renormalization scale $\mu$ of the order of 
 $M_\star\sim|\mbox{Re} \ q_2|$. We have
 \begin{eqnarray}
 q^2_{2} \approx \pm  \frac{1}{G_N N} \frac{120 \pi}{W(-1)} \approx \mp(0.17+0.71\ i)  \frac{120 \pi}{G_N N},
 \end{eqnarray}
 and we thus find the mass of the complex pole:
 \begin{eqnarray}
 m_2=(0.53 -0.67 \ i )  \sqrt{\frac{120 \pi}{G_N N}}.
  \end{eqnarray}
As emphasized before, the mass of this object depends on the number of fields in the theory. The corresponding wave has a frequency:
\begin{eqnarray}
w_2&=&q_2^0=\pm \sqrt{\vec q_2.\vec q_2+(0.17+0.71\ i)  \frac{120 \pi}{G_N N} } 
\\ && \nonumber
= 
\pm \left ( \frac{1}{\sqrt{2}} \sqrt{\sqrt{\left (\vec q_2.\vec q_2+0.17\frac{120 \pi}{G_N N}\right)^2 +\left (0.71 \frac{120 \pi}{G_N N}\right)^2}+
 \vec q_2.\vec q_2+0.17\frac{120 \pi}{G_N N }} \right . \\ &&
 \left . 
  +i   \frac{1}{\sqrt{2}} \sqrt{\sqrt{\left (\vec q_2.\vec q_2+0.17\frac{120 \pi}{G_N N}\right)^2 +\left (0.71 \frac{120 \pi}{G_N N}\right)^2}-
 \vec q_2.\vec q_2-0.17\frac{120 \pi}{G_N N }} \ \right ). \nonumber 
 \end{eqnarray}
 The imaginary part of the complex pole will lead to a damping of the component of the gravitational wave corresponding to that mode.  The complex poles are gravitationally coupled to matter, we must thus assume that the massive modes are produced at the same rate as the usual massless graviton mode if this is allowed kinematically. During an astrophysical event leading to gravitational waves, some of the energy will be emitted into these massive modes which will decay rather quickly because of their large decay width. The possible damping of the gravitational wave implies that care should be taken when relating the energy of the gravitational wave observed on earth to that of the astrophysical event as some of this energy could have been dissipated away as the wave travels towards earth.

The idea that gravitational waves could experience some damping has been considered before \cite{Jones:2015uda}, however it is well known that the graviton cannot split into many gravitons, even at the quantum level \cite{Fiore:1995ai}, if there was such an effect it would have to be at the non-perturbative level \cite{Efroimsky:1994bq}. In our case, the massless mode is not damped, there is thus no contradiction with  the work of \cite{Fiore:1995ai}. Also, as emphasized before the dispersion relation of the massless mode of the gravitational wave is not affected, we do not violate any essentially symmetry such as Lorentz invariance. This is in contrast to the model presented in \cite{Arzano:2016twc}.
  
Since the complex poles couple with the same coupling to matter as the usual massless graviton, we can think of them as a massive graviton although strictly speaking these objects have two polarizations only in contrast to massive gravitons that have five. This  idea has been applied in the context of $F(R)$ gravity \cite{Vainio:2016qas} (see also \cite{Bogdanos:2009tn,Capozziello:2015nga} for earlier works on gravitational waves in $F(R)$ gravity). We shall assume that these massive modes can be excited during the merger of two black holes.  As a rough approximation, we shall assume that all the energy released during the merger is emitted into these modes. Given this assumption, we can use the limit derived by the LIGO collaboration on a graviton mass. We know that $m_g <  1.2 \times 10^{-22}$ eV and we can thus get a limit:
\begin{eqnarray}
 \sqrt{ \mbox{Re}\left (\frac{1}{G_N N} \frac{120 \pi}{ W\left (\frac{-120 \pi M_P^2}{\mu^2 N} \right)} \right) }< 1.2 \times 10^{-22} \ \mbox{eV} 
\end{eqnarray}
we thus obtain a lower bound on $N$: $N>4 \times 10^{102}$ if all the energy of the merger was carried away by massive modes. Clearly this is not realistic as the massless mode will be excited. However, it implies that if the massive modes are produced, they will only arrive on earth if their masses are smaller than $1.2 \times 10^{-22} \ \mbox{eV}$.  Waves corresponding to more massive poles will be damped before reaching earth.  We shall see that there are tighter bounds on the mass of these objects coming from E\"otv\"os type pendulum experiments. 

At this stage, we need to discuss which modes can be produced during the two black holes merger that led to the gravitational wave observed by the LIGO collaboration. The LIGO collaboration estimates that the gravitational wave GW150914 is produced by the coalescence of two black holes: the black holes follow an inspiral orbit before merging, and subsequently going through a final black hole ringdown. Over 0.2 s, the signal increases in frequency and amplitude in about 8 cycles from 35 to 150 Hz, where the amplitude reaches a maximum  \cite{Abbott:2016blz}. The typical energy of the gravitational wave is of the order of 150 Hz or $6\times 10^{-13}$ eV. In other words,  if the gravitational wave had been emitted in the massive mode, they could not have been heavier than $6\times 10^{-22}$ GeV. However, this shows that it is perfectly conceivable that a sizable number of massive gravitons with $m_g <  1.2 \times 10^{-22}$ eV could have been produced. 

Let us now revisit the bound on the number of fields $N$ and thus the new complex pole using E\"otv\"os type pendulum experiments  looking for deviations of the Newtonian $1/r$ potential.   The resummed graviton propagator discussed above can be represented by the effective operator
\begin{eqnarray}
\frac{N}{2304 \pi^2} R \log \left ( \frac{\Box}{\mu^2} \right )R
\end{eqnarray}
where $R$ is the Ricci scalar. As explained above the $\log$ term will be a contribution of order 1, this operator is thus very similar to the more familiar  $ c R^2$ term studied by Stelle long ago. The current bound on the Wilson coefficient of $c$ is $c<10^{61}$ \cite{Hoyle:2004cw,Stelle:1977ry,Calmet:2008tn}. We can translate this bound into a bound on $N$: $N<2\times 10^{65}$. This implies that the mass of the complex pole must be larger than $5\times 10^{-13}$GeV. This bound, although very weak, is more constraining than the one we have obtained from the graviton mass by 37 orders of magnitude.

In this short paper we have investigated quantum gravitational effects in gravitational waves using conservative effective theory methods which are model independent. We found that quantum gravity leads to new poles in the propagator of the graviton besides the usual massless pole. These new states are massive and couple gravitationally to matter. If kinematically allowed, they would thus be produced in roughly the same amount as the usual massless mode in energetic astrophysical events. A sizable amount of the energy produced in astrophysical events could thus be carried away by massive modes which would decay and lead to a damping of this component of the gravitational wave. While our back-of-the-envelope calculation indicates that the energy released in the merger recently observed by LIGO was unlikely to be high enough to produce such modes, one should be careful in extrapolating the amount of energy of astrophysical events from the energy of the gravitational wave observed on earth. This effect could be particularly important for primordial gravitational waves if the scale of inflation is in the region of $10^{16}$ GeV, i.e. within a few orders of magnitude of the Planck scale.


{\it Acknowledgments:}
This work is supported in part by the Science and Technology Facilities Council (grant number  ST/L000504/1), by a Chancellor's International Scholarship of the University of Sussex  and by the National Council for Scientific and Technological Development (CNPq - Brazil).


\bigskip{}

\end{document}